\documentclass[%
superscriptaddress,
preprint,
amsmath,amssymb,
aps,
]{revtex4-1}
\usepackage[T1]{fontenc}
\usepackage[latin9]{inputenc}
\usepackage{amsmath}
\usepackage{amssymb}
\usepackage{stmaryrd}
\usepackage{babel}
\usepackage{hyperref}
\usepackage{rotating}
\usepackage{graphicx}
\usepackage{caption}
\usepackage{subcaption}
\usepackage{float}
\usepackage{nicefrac}
\begin{document}
\title{An intrinsic connection of space-time points}
\author{Ty Shedleski}
\email{tyshedleski@gmail.com}
\affiliation{Alumni, University of Pittsburgh, USA}
\author{Muhammad Usman}
\email{muhammad\_usman\_sharif@yahoo.com}
\email{usman.muhammad@ist.edu.pk}
\affiliation{
	Department of Space Science,
	Institute of Space Technology (IST), Pakistan
}
\affiliation{
	Space and Astrophysics Research Lab (SARL), 
	National Center for GIS and Space Applications (NCGSA), Pakistan
}
\date{\today}
\begin{abstract}
Quantum field theory (QFT) describes the dynamics of quantum particles in the quantum realm in the Minkowski space-time, whereas the General Relativity (GR) is a classical theory describing the nature of dynamical behavior of large bodies in different space-times. This research is a proposal to the proof of concept that through the Einstein-Rosen bridge (also known as wormhole) the information can travel between two points 
thus proving the classical entanglement connection between two spatially distant points which are not causally connected. These results introduce the classical entanglement between the galactic black hole with its surrounding. 
%
%
\end{abstract}
\maketitle
\section{Introduction}\label{introduction}
Quantum Field Theory (QFT) is the concept of explaining the nature and dynamics of the particles at the very basic level depending upon their spin. Spin $0\hbar$ particles are described by the Klein-Gordon equation, spin $\nicefrac{1}{2}\hbar$ particles are described by the Dirac equation and spin $1\hbar$ are described by the Proca equation. The hypothetical spin $2\hbar$ particles i.e. graviton is yet to be completely understood. On the other side (of relativity), the dynamics of any object or system of objects can be described using the Einstein Field Equations (EFEs) but its consequences and implications on the level of black holes, galaxies and cosmology are yet to be completely understood as well as checked observationally. 

In this article, we demonstrate the energy flow from a harmonic oscillator the wormhole. The results enable us to understand the concept of the wormhole which is the prediction of the general relativity. Wormholes mathematically come as the solution of the EFEs that are the (hypothetical) structures of space-time that connect the causally disconnected space-time points. We find that the energy flow will happen when a harmonic oscillator is placed in the space-time of a wormhole and the phase-space is not restricted. The article's structure is as follows: in section (\ref{Sec:KG-eq-in-Curved-Spacetime}), we describe the Klein-Gordon equation in the curved space-time, section (\ref{Sec:Dirac-Equation}) describes the Dirac equation in the curved space-time. In section (\ref{Sec:ER-Bridge}), the dynamics of wormhole, harmonic oscillator when placed in the space-time of wormhole and energy flow is presented. Section (\ref{conclusion}) concludes the article with summery of the work.
\section{The Klein-Gordon Equation in Curved Space-Time} \label{Sec:KG-eq-in-Curved-Spacetime}
We start this section by writing the Klein-Gordon Lagrangian density in flat space-time 
\begin{equation}\label{eq:Klein-Gordon-Lagrangian}
	\mathcal{L}=\frac{1}{2}\eta^{\mu\nu}\left(\partial_{\nu}\phi\right)\left(\partial_{\mu}\phi\right)-\frac{1}{2}m^{2}\phi^{2}
\end{equation}
$\phi$ is a scalar, therefore its covariant derivative will be equivalent to the partial derivative $\partial_{\nu}\phi$. The space-time curvature will lead to the addition of another term in eq. (\ref{eq:Klein-Gordon-Lagrangian}), proportional to the Ricci curvature $R.$ Therefore, the Klein-Gordon Lagrangian density in curved space-time can be written in a general form 
\begin{align}
	\mathcal{L} & =\mathcal{L}_{flat}+\mathcal{L}_{curved}\nonumber \\
	& =\frac{1}{2}\eta^{\mu\nu}\left(\partial_{\nu}\phi\right)\left(\partial_{\mu}\phi\right)-\frac{1}{2}m^{2}\phi^{2}-\frac{1}{2}\xi R\phi^{2} \label{eq:Klein-Gordon-Lagrangian-curved}
\end{align}
where $\xi$ is the gravitational coupling. Let us recall the action in curved space-time and the d'Alembert rule
\begin{align*}
	S & =\int d^{4}x\sqrt{-g}\mathcal{L}\\
	\delta S & =0
\end{align*}
the solution of the d'Alembert rule is the Euler-Lagrange equations
\begin{align}
	\delta S & =\partial_{\mu}\left(\dfrac{\partial\left(\sqrt{-g}\mathcal{L}\right)}{\partial\left(\partial_{\mu}\phi\right)}\right)-\dfrac{\partial\left(\sqrt{-g}\mathcal{L}\right)}{\partial\phi}=0\nonumber
\end{align}
Thus
\begin{equation}\label{eq:Euler-Lagrange-equation}
	\partial_{\mu}\left(\dfrac{\partial\left(\sqrt{-g}\mathcal{L}\right)}{\partial\left(\partial_{\mu}\phi\right)}\right)-\dfrac{\partial\left(\sqrt{-g}\mathcal{L}\right)}{\partial\phi}=0
\end{equation}
by inserting eq. (\ref{eq:Klein-Gordon-Lagrangian-curved}) into Euler-Lagrange equations, we obtain the Klein-Gordon equation in curved space-time 
\begin{equation}\label{eq:Klein-Gordon-equation-curved-space-time}
	\frac{1}{\sqrt{-g}}\partial_{\mu}\left(\sqrt{-g}g^{\mu\nu}\partial_{\nu}\phi\right)+\left(m^{2}+\xi R\right)\phi=0
\end{equation}
As an example, for the FRW metric $g$, we write
\begin{equation}\label{eq:sqrt-metric}
	\sqrt{-g}=\frac{a^{3}r^{2}\sin\theta}{\sqrt{1-kr^{2}}}
\end{equation}
the eq. (\ref{eq:Klein-Gordon-equation-curved-space-time}) can be rearranged 
\begin{align}
	\frac{1}{\sqrt{-g}}\partial_{\mu}\left(\sqrt{-g}g^{\mu\nu}\partial_{\nu}\phi\right)+\left(m^{2}+\xi R\right)\phi & =0\nonumber \\
	\frac{1}{\sqrt{-g}}\partial_{0}\left(\sqrt{-g}g^{00}\partial_{0}\phi\right)+\frac{1}{\sqrt{-g}}\partial_{i}\left(\sqrt{-g}g^{ij}\partial_{j}\phi\right)+\left(m^{2}+\xi R\right)\phi & =0\nonumber \\
	\partial_{0}^{2}\phi+\frac{1}{\sqrt{-g}}\partial_{i}\left(\sqrt{-g}g^{ij}\partial_{j}\phi\right)+\left(m^{2}+\xi R\right)\phi & =0
\end{align}
where the 0 indice is for the temporal component of the metric and $i,j=1,3$ are the spatial components. Note that there are only three equations instead of four because $\theta\theta$ and $\phi\phi$ equations are the same. The Laplacian for FRW space-time is 
\begin{equation}\label{eq:Laplacian-spherical-coordinates}
	\nabla=\frac{\sqrt{1-kr^{2}}}{r^{2}}\frac{\partial}{\partial r}\left(\sqrt{1-kr^{2}}r^{2}\frac{\partial}{\partial r}\right)+\frac{1}{r^{2}\sin^{2}\theta}\frac{\partial}{\partial\theta}\left(\sin\theta\frac{\partial}{\partial\theta}\right)+\frac{1}{r^{2}\sin^{2}\theta}\left(\frac{\partial^{2}}{\partial\varphi^{2}}\right)
\end{equation}
\section{Dirac Equation in Curved Space-Time}\label{Sec:Dirac-Equation}
Let us also discuss the Dirac equation in curved space-time, i.e. the covariant derivative and the coordinates are computed in a Riemannian Manifold. Recall that the general Dirac equation is given by 
\begin{equation}\label{Eq:Dirac-Equation}
\left(i\gamma^{\mu}\nabla_{\mu}-m\right)\psi=0~,
\end{equation}
the covariant derivative in the above equation $\nabla_{\mu}$ is equivalent to $\partial_{\mu}$ in a Minkowskian manifold (flat space-time). 
The covariant derivative is defined as 
\begin{equation}\label{Eq:covariant-derivative-defination}
\nabla_{\mu}\equiv\partial_{\mu}+\Gamma_{\mu}~,
\end{equation}
where $\Gamma_{\mu}$ are the spin connections also known as the Christoffel's symbols (for torsion free space). They are given as
\begin{equation}\label{Eq:Christoffel's-symbols}
\Gamma_{\mu}=\frac{1}{2}\sigma^{bc}e_{b}^{\nu}e_{c\nu;\mu}~.
\end{equation}
It is important to discuss the appropriate Dirac matrices, $\gamma^{\mu}$, in the given curved space-time. The metric tensor of the curved space-time is defined in a general coordinate system by
\begin{equation}\label{Eq:metric}
g_{\mu\nu}\left(x\right)=e_{\mu}^{a}\left(x\right)e_{\nu}^{b}\left(x\right)\eta_{ab}
\end{equation}
where $e_{\mu}^{a}$, known as the dyads or the vielbein or simply the vector directions, are 
\begin{equation}\label{Eq:vector-directions}
e_{\mu}^{a}\left(x\right)=\left(\frac{\partial Y_{X}^{a}}{\partial x^{\mu}}\right)_{x=X},\qquad a=0,\,1
\end{equation}
therefore, the $\gamma$ matrices in curved space-time are defined over the dyads by 
\begin{equation}\label{Eq:gamma}
\gamma^{\mu}=e_{\mu}^{a}\gamma^{a}
\end{equation}
in the eq. (\ref{Eq:gamma}), the $\gamma^{a}$ are the flat Mikowskian $\gamma$
matrices which satisfy 
\begin{equation}\label{Eq:Gamma-matrices-relation}
\left\{ \gamma^{a},\gamma^{b}\right\} =2\eta^{ab}~,\qquad\frac{1}{4}\left[\gamma^{a},\gamma^{b}\right]=\sigma^{ab}~.
\end{equation}
Using eq. (\ref{Eq:metric}) and eq. (\ref{Eq:gamma}), a straightforward calculation gives us the definition of the $\gamma$ matrices in curved space-time 
\begin{equation}\label{eq:Gamma-relation}
\left\{ \gamma^{\mu},\gamma^{\nu}\right\} =2g^{\mu\nu}
\end{equation}
The importance of equation \ref{Eq:Gamma-matrices-relation} is in the definition of the commutation rule when used in the covariant derivative commutation relation with the gamma matrices
\begin{equation}\label{eq:commutation-Gamma-gamma}
	\left[\Gamma_{\mu},\gamma^{\mu}\right]=\frac{\partial\gamma^{\mu}}{\partial x^{\mu}}+\Gamma^{\mu}_{\mu\rho}\gamma^{\rho},
\end{equation}
that leads us to introduce the Christoffel symbols 
\begin{align}
\Gamma_{\nu\rho}^{\mu} & =\frac{1}{2}g^{\lambda\mu}\left[g_{\lambda\nu,\rho}+g_{\lambda\rho,\nu}-g_{\nu\rho,\lambda}\right]\nonumber \\[1ex]
\Gamma_{\mu\rho}^{\mu}&=\frac{1}{\sqrt{-g}}\partial_{\rho}\left(\sqrt{-g}\right)~. \label{eq:Gamma-mu=nu}
\end{align}
For simplicity, considering the simple (1+1) space-time case, eq. (\ref{eq:commutation-Gamma-gamma}) gives 
\begin{equation}
\gamma^{\mu}\Gamma_{\mu}=\frac{1}{2}\gamma^{a}\frac{1}{\sqrt{-g}}\partial_{\mu}\left(\sqrt{-g}e_{a}^{\mu}\right)
\end{equation}
Then the Dirac equation given by eq. (\ref{Eq:Dirac-Equation}) now can be written 
\begin{align*}
&\left(i\gamma^{\mu}\nabla_{\mu}-m\right)\psi =0\\
&\left[i\gamma^{\mu}\left(\partial_{\mu}+\Gamma_{\mu}\right)-m\right]\psi =0\\
&\left[i\gamma^{\mu}\partial_{\mu}+i\gamma^{\mu}\Gamma_{\mu}-m\right]\psi =0\\
&\left[ie_{a}^{\mu}\gamma^{a}\partial_{\mu}+i\frac{1}{2}\gamma^{a}\frac{1}{\sqrt{-g}}\partial_{\mu}\left(\sqrt{-g}e_{a}^{\mu}\right)-m\right]\psi =0
\end{align*}
The Dirac equation, in natural units, in curved space-time is 
\begin{equation}
\left[i\gamma^{a}e_{a}^{\mu}\partial_{\mu}+\frac{i}{2}\gamma^{a}\frac{1}{\sqrt{-g}}\partial_{\mu}\left(\sqrt{-g}e_{a}^{\mu}\right)-m\right]\psi=0
\end{equation}
Let us now focus on the space-time part, since we are considering the work in (1+1) space-time, a special feature of this space-time is that the metric can always be reduced to the conformally flat form 
\begin{equation}
ds^{2}=\Omega^{2}\left(dt^{2}-dx^{2}\right)
\end{equation}
the function $\Omega\left(x,t\right)$ is pre-defined folowing the chosen metric. We here are more interested at this moment in the general behavior of this function to derive the Dirac equation. The Christoffel symbols are given by 
\begin{equation}
\begin{cases}
\Gamma_{00}^{0}=\Gamma_{11}^{0}=\Gamma_{10}^{1}=\Gamma_{01}^{1}=\frac{\dot{\Omega}}{\Omega}\\
\Gamma_{01}^{0}=\Gamma_{10}^{0}=\Gamma_{00}^{1}=\Gamma_{11}^{1}=\frac{\Omega'}{\Omega}
\end{cases}
\end{equation}
The 0 and 1 stands for the time and space coordinates respectively,
while $\dot{\Omega}=\frac{\partial\Omega}{\partial x^{0}}=\frac{\partial\Omega}{\partial t}$
and $\Omega'=\frac{\partial\Omega}{\partial x^{1}}=\frac{\partial\Omega}{\partial x}$.
The vielbein $e_{a}^{\mu}\left(x\right)$ are given by 
\begin{equation}\label{eq:vielbein}
e_{a}^{\mu}\left(x\right)=\left.\frac{\partial x^{\mu}}{\partial y^{a}}\right|_{x^{\mu}=X^{\mu}}\rightarrow\left\{ e_{a}^{\mu}=\frac{1}{\Omega}\right\} 
\end{equation}
Another way to write the spin connections is 
\begin{equation}\label{eq:spin-connections}
\Gamma_{\mu}\left(x\right)=-\frac{i}{4}\omega_{ab\nu}\left(x\right)\sigma^{ab}
\end{equation}
where $\sigma^{ab}=\frac{i}{2}\left[\gamma^{a},\gamma^{b}\right]$. It is now straightforward to show $\left[\gamma^{a},\Gamma_{\mu}\right]=\omega_{b\nu}^{a}\gamma^{b}$. Thus we can easily obtain the spin connection in terms of the affine connection 
\begin{equation}\label{eq:affine-connection}
\omega_{b\nu}^{a}=e_{\mu}^{a}\partial_{\nu}\left(e_{b}^{\mu}\right)+e_{\mu}^{a}e_{b}^{\sigma}\Gamma_{\sigma\nu}^{\mu}
\end{equation}
Now we compute the spin connection on our (1+1) spacetime 
\begin{align*}
\omega_{10}^{0} & =e_{\mu}^{0}\partial_{0}\left(e_{1}^{\mu}\right)+e_{\mu}^{0}e_{1}^{\sigma}\Gamma_{\sigma0}^{\mu}\\[1ex]
 &=e_{1}^{0}\partial_{0}\left(e_{1}^{1}\right)+e_{0}^{0}\partial_{0}\left(e_{1}^{0}\right)e_{1}^{0}e_{1}^{0}\Gamma_{00}^{1}+e_{0}^{0}e_{1}^{0}\Gamma_{00}^{0}+e_{1}^{0}e_{1}^{1}\Gamma_{10}^{1}+e_{0}^{0}e_{1}^{1}\Gamma_{10}^{0}\\[1ex]
& =\frac{\Omega'}{\Omega} \\[2ex] 
\omega_{11}^{0} & =e_{\mu}^{0}\partial_{0}\left(e_{1}^{\mu}\right)+e_{\mu}^{0}e_{1}^{\sigma}\Gamma_{\sigma1}^{\mu}\\[1ex]
& =e_{1}^{0}\partial_{0}\left(e_{1}^{1}\right)+e_{0}^{0}\partial_{0}\left(e_{1}^{0}\right)+e_{1}^{0}e_{1}^{0}\Gamma_{01}^{1}+e_{0}^{0}e_{1}^{0}\Gamma_{01}^{0}+e_{1}^{0}e_{1}^{1}\Gamma_{11}^{1}+e_{0}^{0}e_{1}^{1}\Gamma_{11}^{0}\\[1ex]
& =\frac{\dot{\Omega}}{\Omega}
\end{align*}
Thus using symmetry
\begin{equation}\label{eq:omega}
\begin{cases}
\omega_{10}^{0}=\omega_{00}^{1}=\frac{\Omega'}{\Omega}\\
\omega_{11}^{0}=\omega_{01}^{1}=\frac{\dot{\Omega}}{\Omega}
\end{cases}
\end{equation}
The equation (\ref{eq:spin-connections}) now leads to a general definition of $\Gamma_{\mu}$ for
$\mu=0,\,1$ 
\begin{align}
\Gamma_{0} & =-\frac{i}{4}\omega_{010}\frac{i}{2}\left[\gamma^{0},\gamma^{1}\right]\nonumber \\
 & =\frac{\Omega'}{8\Omega}\left[\gamma^{0},\gamma^{1}\right] \label{eq:Gamma0}
\end{align}
\vspace{-0.7cm}
\begin{align}
\Gamma_{1} & =-\frac{i}{4}\omega_{101}\frac{i}{2}\left[\gamma^{0},\gamma^{1}\right]\nonumber \\
 & =\frac{\dot{\Omega}}{8\Omega}\left[\gamma^{0},\gamma^{1}\right]\label{eq:Gamma1}
\end{align}
\textbf{Remarks:} 
Recall the Dirac equation given by eq. (\ref{Eq:Dirac-Equation}) 
\begin{align}
&\left(i\gamma^{\mu}\nabla_{\mu}-m\right)\psi\left(x\right) =0\nonumber \\
&\left[i\gamma^{\mu}\partial_{\mu}+i\gamma^{\mu}\Gamma_{\mu}-m\right]\psi\left(x\right) =0\nonumber \\
&\left[i\gamma^{0}\partial_{0}+i\gamma^{1}\partial_{1}+i\gamma^{0}\Gamma_{0}+i\gamma^{1}\Gamma_{1}-m\right]\psi\left(x\right) =0\nonumber \\
&\left[i\gamma^{0}\partial_{0}+i\gamma^{1}\partial_{1}+i\gamma^{0}\frac{\Omega'}{8\Omega}\left[\gamma^{0},\gamma^{1}\right]+i\gamma^{1}\frac{\dot{\Omega}}{8\Omega}\left[\gamma^{0},\gamma^{1}\right]-m\right]\psi\left(x\right) =0 \label{eq:Dirac-equation-new}
\end{align}
multiplying 
by $\Omega\gamma^{0}$ and using $\gamma^{\mu}=e_{\mu}^{a}\gamma^{a}$
\begin{align*}
&\left[i\Omega\gamma^{0}\gamma^{0}\partial_{0}+i\Omega\gamma^{0}\gamma^{1}\partial_{1}+i\Omega\gamma^{0}\gamma^{0}\frac{\Omega'}{8\Omega}\left[\gamma^{0},\gamma^{1}\right]+i\Omega\gamma^{0}\gamma^{1}\frac{\dot{\Omega}}{8\Omega}\left[\gamma^{0},\gamma^{1}\right]-\Omega\gamma^{0}m\right]\psi\left(x\right)=0\\
& \left[i\partial_{0}+i\gamma^{0}\gamma^{1}\partial_{1}+i\frac{\Omega'}{8\Omega}\left[\gamma^{0},\gamma^{1}\right]+i\gamma^{0}\gamma^{1}\frac{\dot{\Omega}}{8\Omega}\left[\gamma^{0},\gamma^{1}\right]-\Omega\gamma^{0}m\right]\psi\left(x\right)=0
\end{align*}
With some trivial rearrangement, we obtain 
\begin{equation}\label{eq:Dirac-equation-new-new}
i\left[\partial_{0}+\gamma^{0}\gamma^{1}\frac{\left[\gamma^{0},\gamma^{1}\right]}{8}\frac{\dot{\Omega}}{\Omega}\right]\psi\left(x\right)=i\left[\gamma^{0}\gamma^{1}\partial_{1}+\frac{\left[\gamma^{0},\gamma^{1}\right]}{8}\frac{\Omega'}{\Omega}\right]\psi\left(x\right)+\Omega\gamma^{0}m\,\psi\left(x\right)
\end{equation}
If we choose $\gamma^{0}=\sigma_{z}$ and $\gamma^{1}=i\sigma_{y},$where
$\sigma_{i}$ are the Pauli matrices. The eq. (\ref{eq:Dirac-equation-new-new}) can be rewritten as
\begin{equation}
i\left(\partial_{0}+\frac{\dot{\Omega}}{4\Omega}\right)\psi\left(x\right)=\left[-i\sigma_{x}\left(\partial_{1}+\frac{\Omega'}{4\Omega}\right)+\Omega\sigma_{z}m\right]\psi\left(x\right)
\end{equation}
which completely is in agreement with the Dirac equation in flat space-time,
i.e $\Omega=1$, 
\begin{equation}\label{eq:Dirac-equation-flat-space-time}
i\partial_{0}\psi\left(x\right)=-i\sigma_{x}\partial_{1}\psi\left(x\right)+\sigma_{z}m\psi\left(x\right)
\end{equation}
\section{The Einstein-Rosen Bridge, Wormhole and Energy Flow}\label{Sec:ER-Bridge}
Time travel in general relativity is described by the theories of wormhole.
Consider the charge free Schwarzschild space-time
\begin{equation}\label{eq:Schwarzschild-spacetime}
	\mathrm{d}s^2 = -c^2\left(1-\dfrac{2Gm}{r c^2}\right)\mathrm{d}t^2+\dfrac{1}{1-\dfrac{2Gm}{r c^2}}\mathrm{d}r^2+r^2\left(\mathrm{d\theta^2+\sin^2\theta\mathrm{d}\phi^2}\right)~.
\end{equation}
Using the transformation 
$$u^2=c^2r-2Gm~,$$ 
the Schwarzschild space-time becomes
\begin{equation}\label{eq:Schwarzschild-spacetime-transformed}
	\mathrm{d}s^2 = -c^2\dfrac{u^2}{u^2+2Gm}\mathrm{d}t^2+4\dfrac{u^2+2Gm}{c^4}\mathrm{d}u^2+\left(\dfrac{u^2+2Gm}{c^2}\right)^2\left(\mathrm{d\theta^2+\sin^2\theta\mathrm{d}\phi^2}\right)~.
\end{equation}
Evidently, the above space-time can be described by two different congruent parts or sheets, one corresponding to $u<0$ while the other corresponding to $u>0$. These two sheets of planes are connected via a hyperplane $u=0$ (corresponding to $r=2Gm/c^2$) in which $g$ vanishes. This connection between the two sheets is the Einstein-Rosen bridge or wormhole which is not restricted by the gravity on either side.

The inclusion of electric charge in the Schwarzschild space-time leads to the following Schwarzschild static spherically symmetric space-time
\begin{equation}\label{eq:Schwarzschild-charged-spacetime}
	\mathrm{d}s^2 = -c^2\left(1-\dfrac{2Gm}{r c^2}-\dfrac{G}{4\pi\epsilon_0 c^4}\dfrac{\varepsilon^2}{2r^2}\right)\mathrm{d}t^2+\dfrac{1}{1-\dfrac{2Gm}{r c^2}-\dfrac{G}{4\pi\epsilon_0 c^4}\dfrac{\varepsilon^2}{2r^2}}\mathrm{d}r^2+r^2\left(\mathrm{d\theta^2+\sin^2\theta\mathrm{d}\phi^2}\right)~.
\end{equation}
where $\varepsilon$ is the electric charge. Using the transformation with $m=0$ $$u^2=\dfrac{1}{2}\left(8\pi\epsilon_0 c^4r^2-G\varepsilon^2\right)~,$$ 
the eq. (\ref{eq:Schwarzschild-charged-spacetime}) becomes
\begin{equation}\label{eq:Schwarzschild-charged-spacetime-transformed}
	\mathrm{d}s^2 = -c^2\dfrac{2u^2}{\left(2u^2+G\varepsilon^2\right)}\mathrm{d}t^2+\mathrm{d}u^2+\left(\dfrac{2u^2+G\varepsilon^2}{8\pi\epsilon_0 c^4}\right)\left(\mathrm{d\theta^2+\sin^2\theta\mathrm{d}\phi^2}\right)~.
\end{equation}
Once again, quite clear that the above space-time can be described by two different congruent parts or sheets, one corresponding to $u<0$ while the other corresponding to $u>0$. These two sheets of planes are connected via a hyperplane $u=0$ (corresponding to $r^2=G\varepsilon^2/8\pi\epsilon_0 c^4$) in which $g$ vanishes. This connection between the two sheets is the Einstein-Rosen bridge or wormhole which is not restricted by the gravity on either side.

The neutral space-time interval corresponding to the non-traversable wormhole is given by eq. (\ref{eq:Schwarzschild-spacetime-transformed}). The phase-space of the wormhole allows for the energy flow through the wormhole neck. Imagine a situation in which an harmonic oscillator is oscillating. In this scenario, we want to see how the oscillation in the oscillator will be effected by the presence of the wormhole. In this case, the Einstein File Equations (EFEs) are needed to be solved in the corresponding space-time with matter component taken as the harmonic oscillator.

The Einstein tensor corresponding to the space-time given by eq. (\ref{eq:Schwarzschild-spacetime-transformed}) is 
\begin{eqnarray}\label{eq:Einstein-Tensor}
	G_{tt}&=&\dfrac{u^2\left(c^4+16G m u'^2+8 u^2 u'^2\right)}{\left(2G m +u^2\right)^3} \\
	G_{rr}&=&\dfrac{4\left(-c^6 r + 8 \left(-2 G m u - u^3\right)u'^2-8\left(2G m + u^2\right)u''\right)}{c^6\left(2G m u + u^3\right)} \\
	G_{\theta\theta}&=&\dfrac{-4 u \left(2 G m + u^2\right)u'^2-6\left(2 G m + u^2\right)u''}{c^6 u} \\
	G_{\phi\phi}&=&\dfrac{\left(-4u\left(2 G m + u^2\right)u'^2-6\left(2 G m + u^2\right)^2u''\right)\sin(\theta)^2}{c^6 u} \\
	G_{\mu\nu}&=&0 ;\qquad\qquad\qquad\qquad\qquad\qquad \text{when }\mu\neq\nu
\end{eqnarray}
The stress-energy tensor for the harmonic oscillator is given by 
\begin{eqnarray}\label{eq:Stress-energy-Tensor}
	T_{tt}&=&-\left(\dfrac{1}{2}Mr'^2+\dfrac{1}{2}M\omega^2r^2\right) \\
	T_{rr}&=& \dfrac{1}{2}Mr'^2-\dfrac{1}{2}M\omega^2r^2 \\
	T_{\theta\theta}&=&\dfrac{1}{2}Mr'^2-\dfrac{1}{2}M\omega^2r^2 \\
	T_{\phi\phi}&=&\dfrac{1}{2}Mr'^2-\dfrac{1}{2}M\omega^2r^2 \\
	T_{\mu\nu}&=&0 ;\qquad\qquad\qquad\qquad\qquad \text{when }\mu\neq\nu
\end{eqnarray}
%

Equating both sides gives the Einstein Field Equations. Also the ${\theta\theta}$ and ${\phi\phi}$ components are the same so only one will be kept. The solution of the Einstein field equations gives the dynamics of harmonic oscillator placed in the wormhole. These are
\begin{eqnarray}\label{eq:EFEs}
	\dfrac{u^2\left(c^4+16G m u'^2+8 u^2 u'^2\right)}{\left(2G m +u^2\right)^3}&=&-\dfrac{8\pi G}{c^4}\left(\dfrac{1}{2}Mr'^2+\dfrac{1}{2}M\omega^2r^2\right) \label{EFE-1} \\
	\dfrac{4\left(-c^6 r + 8 \left(-2 G m u - u^3\right)u'^2-8\left(2G m + u^2\right)u''\right)}{c^6\left(2G m u + u^3\right)}&=&\dfrac{8\pi G}{c^4} \left(\dfrac{1}{2}Mr'^2-\dfrac{1}{2}M\omega^2r^2\right) \label{EFE2}\\
	\dfrac{-4 u \left(2 G m + u^2\right)u'^2-6\left(2 G m + u^2\right)u''}{c^6 u}&=&\dfrac{8\pi G}{c^4} \left(\dfrac{1}{2}Mr'^2-\dfrac{1}{2}M\omega^2r^2\right) \label{EFE3}
\end{eqnarray}
where $m$ is the mass of the black hole describing the geometry given by eq. (\ref{eq:Schwarzschild-spacetime-transformed}), $M$ is the mass of the harmonic oscillator placed in the space-time given by eq. (\ref{eq:Schwarzschild-spacetime-transformed}), $u(t)$ is the transformed spatial parameter of the black hole geometry, $r(t)$ is the radial parameter of the harmonic oscillator and $\omega$ is the angular frequency of the harmonic oscillator. From Eq. (\ref{EFE-1}), $M$ can be evaluated as 
\begin{equation}\label{EFE1-M}
	M=-\dfrac{c^4}{4\pi G}\left(\dfrac{1}{r'^2+\omega^2 r^2}\right)\left(\dfrac{u^2\left(c^4+16G m u'^2+8 u^2 u'^2\right)}{\left(2G m +u^2\right)^3}\right)
\end{equation}
Using the above equation in the remaining two equations (eq. (\ref{EFE2})and (\ref{EFE3})) and then solving the obtained equations numerically to get the dynamics of the harmonic oscillator placed in the wormhole. The initial conditions used are $u(0)=1$, $u'(0)=0$ and $r(0)=1$. We have also used the natural units so that the solution is simplified. The initial condition of harmonic oscillator radial parameter $r(0)=1$ represents the perturbation at the starting time $t=0$ while the initial conditions of the radial parameter of the wormhole represent placement of oscillator is happening and due to the perturbation the space-time of the wormhole is being disturbed equally. We obtained two sets of solutions of the eqs. (\ref{EFE-1})-(\ref{EFE3}) which are
\subsection{Solution-I}\label{Solution-I}
\begin{figure}[h!]
	\centering
	\includegraphics[width=0.7\linewidth]{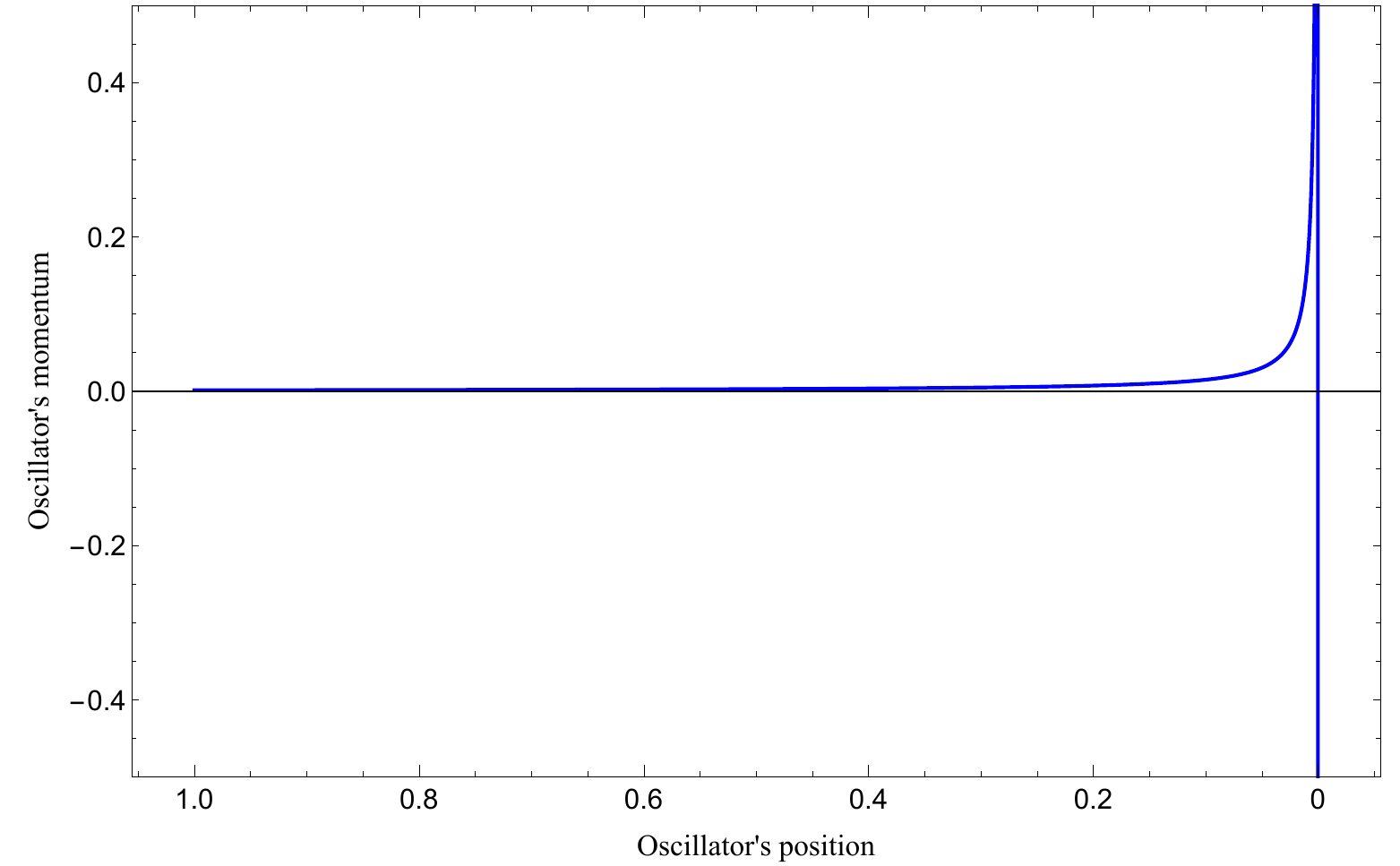}
	\caption{Phase-space diagram of the oscillator.}
	\label{fig:um-oscillator-phasespace-solution-1}
\end{figure}

In this outcome scenario, the dynamics of the oscillators happens such that the velocity of the oscillator asymptotically increases such that the momentum shoots to infinity as the oscillator approaches the event horizon. This is shown in fig. (\ref{fig:um-oscillator-phasespace-solution-1}).
\begin{figure}[h!]
	\centering
	\begin{subfigure}[b]{0.7\textwidth}
		\includegraphics[width=1\textwidth]{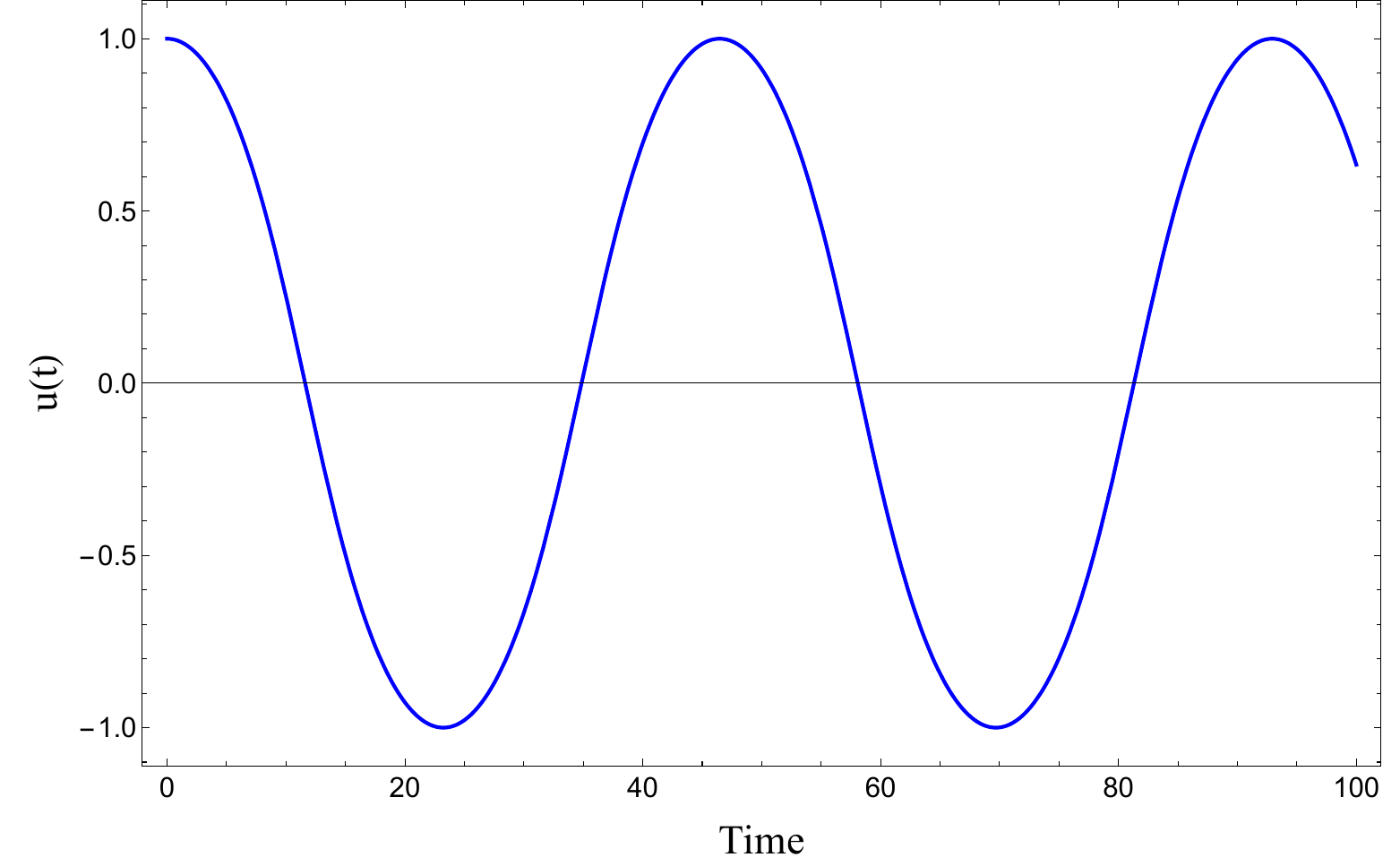}
		\caption{Evolution of radial parameter of wormhole.}
		\label{fig:um-BH-solution-1}
	\end{subfigure}%
	~ 
	~ 
	\caption{wormhole dynamics.}
	\label{fig:wormhole-dynamics.}
\end{figure}

From the plot of the fig. (\ref{fig:wormhole-dynamics.}), in which the radial parameter of the wormhole is plotted as the function of time, we see that the wormhole does allow for the oscillatory behavior to occur in an ideal scenario. There is once again no damping effect in the ideal oscillatory behavior suggesting that the energy can ideally flow without any loss (the case when there is some object placed in the space-time is discussed in the next paragraph). 

\begin{figure}[h!]
	\centering
	\begin{subfigure}[b]{0.5\textwidth}
		\includegraphics[width=\textwidth]{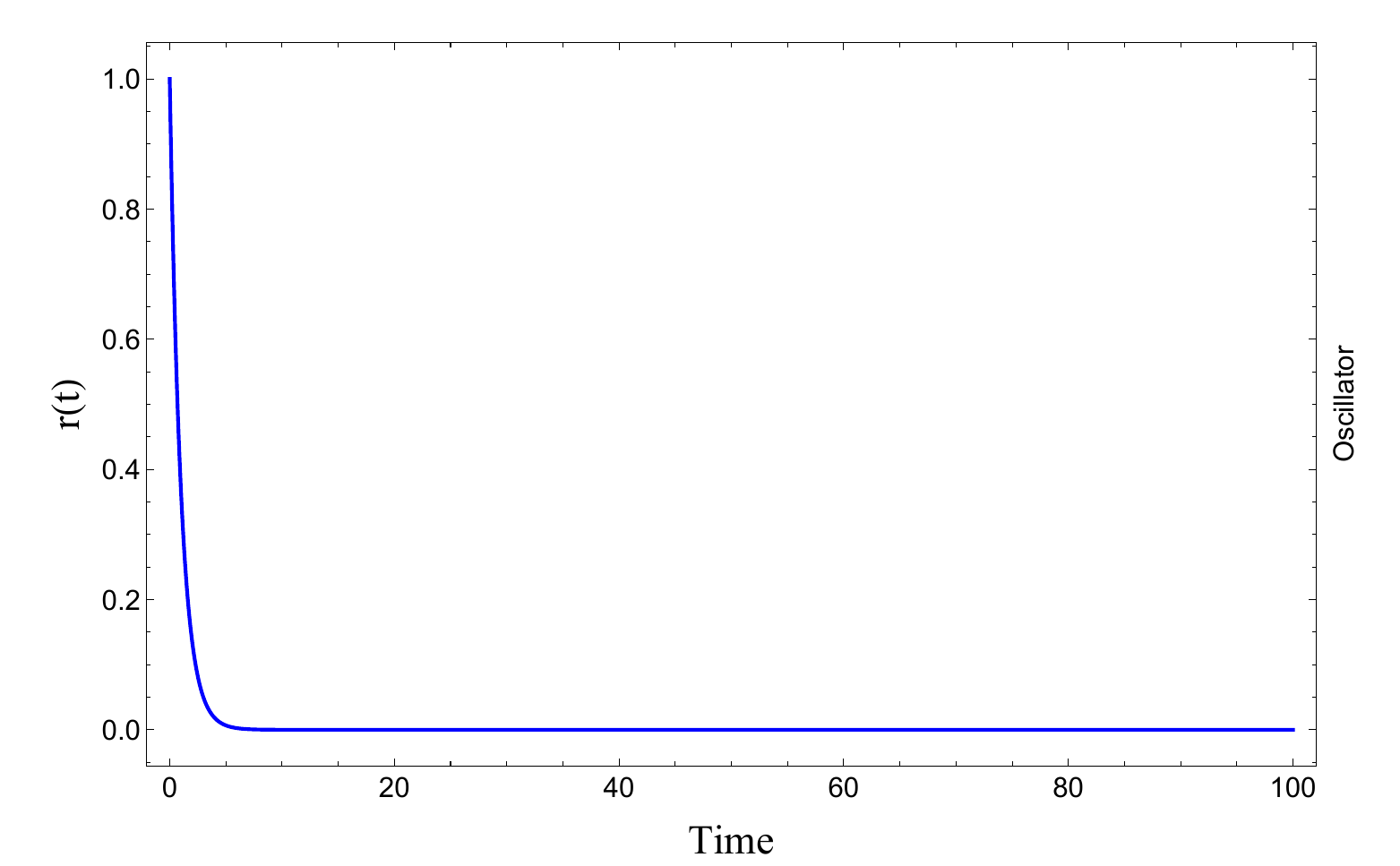}
		\caption{Evolution of oscillator's radial parameter. \qquad\qquad}
		\label{fig:um-oscillator-solution-1}
	\end{subfigure}%
	~ 
	\begin{subfigure}[b]{0.5\textwidth}
		\includegraphics[width=\textwidth]{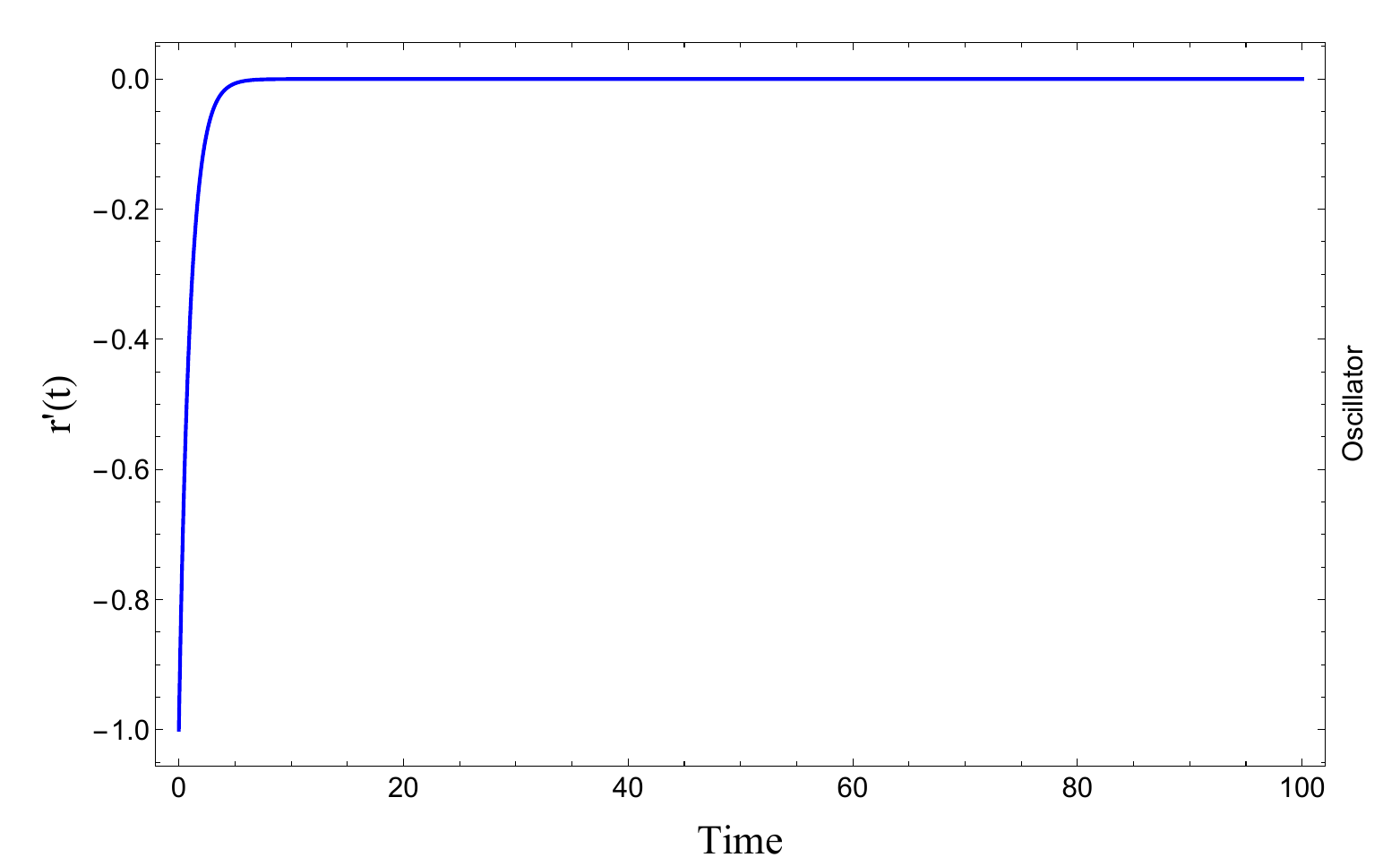}
		\caption{Evolution of oscillator's radial parameter's derivative.}
		\label{fig:uM-Oscillator-prime-solution-1}
	\end{subfigure}
	~ 
	\caption{Harmonic oscillator's dynamics.}\label{fig:Harmonic-oscillator's-dynamics.}
\end{figure}

The plots shown in fig. (\ref{fig:Harmonic-oscillator's-dynamics.}) are the dynamics of harmonic oscillator when placed in the wormhole space-time. Here we see that as we place an oscillator in the region, the gravity of the massive object (Schwarzschild black hole here) causes the oscillator to lose energy. The damping of the oscillations are ultra high and the oscillations completely vanish in the first damping. The initial energy $E_0$ of the oscillator is trasferred to the space-time which would move through the wormhole.

The phase-space diagram of this scenario is presented in fig. (\ref{fig:um-oscillator-phasespace-solution-1}). As explained earlier, the dynamics of the oscillators happens such that the velocity of the oscillator asymptotically increases such that the momentum shoots to infinity as the oscillator approaches the event horizon. $0$ here represents the location of the event horizon. This gain in the momentum would come from the transfer of energy from the other side of the wormhole. 
\subsection{Solution-II}\label{Solution-II}
\begin{figure}[h!]
	\centering
	\includegraphics[width=0.7\linewidth]{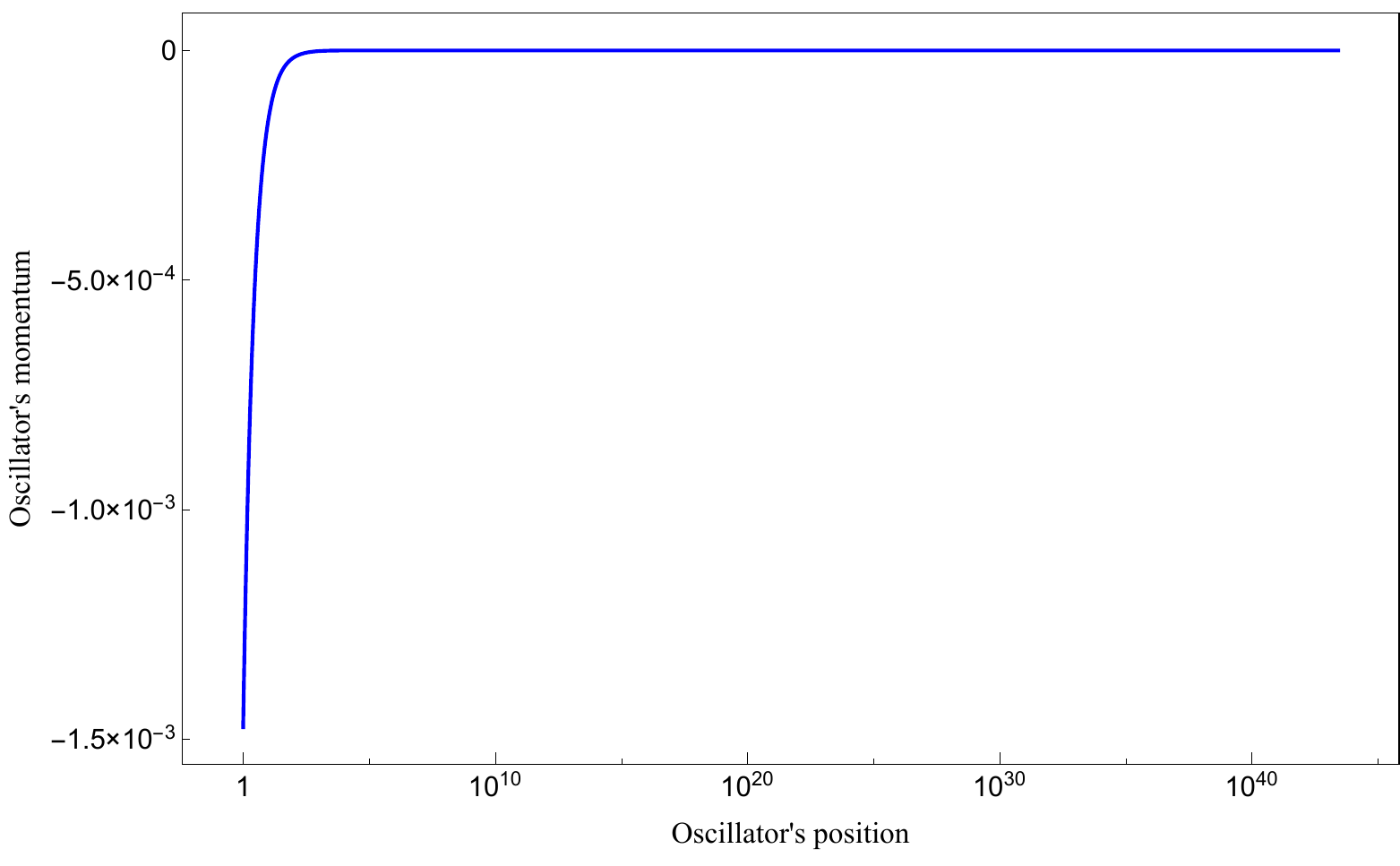}
	\caption{Phase-space diagram of the oscillator.}
	\label{fig:um-oscillator-phasespace-solution-2}
\end{figure}

In the second outcome scenario, the dynamics of the oscillators happens in such a way that the velocity of the oscillator asymptotically increases such that the momentum  in this scenario does not shoot to infinity as the oscillator approaches the event horizon. This is shown in fig. (\ref{fig:um-oscillator-phasespace-solution-2}). 
\begin{figure}[h!]
	\centering
	\begin{subfigure}[b]{0.7\textwidth}
		\includegraphics[width=1\textwidth]{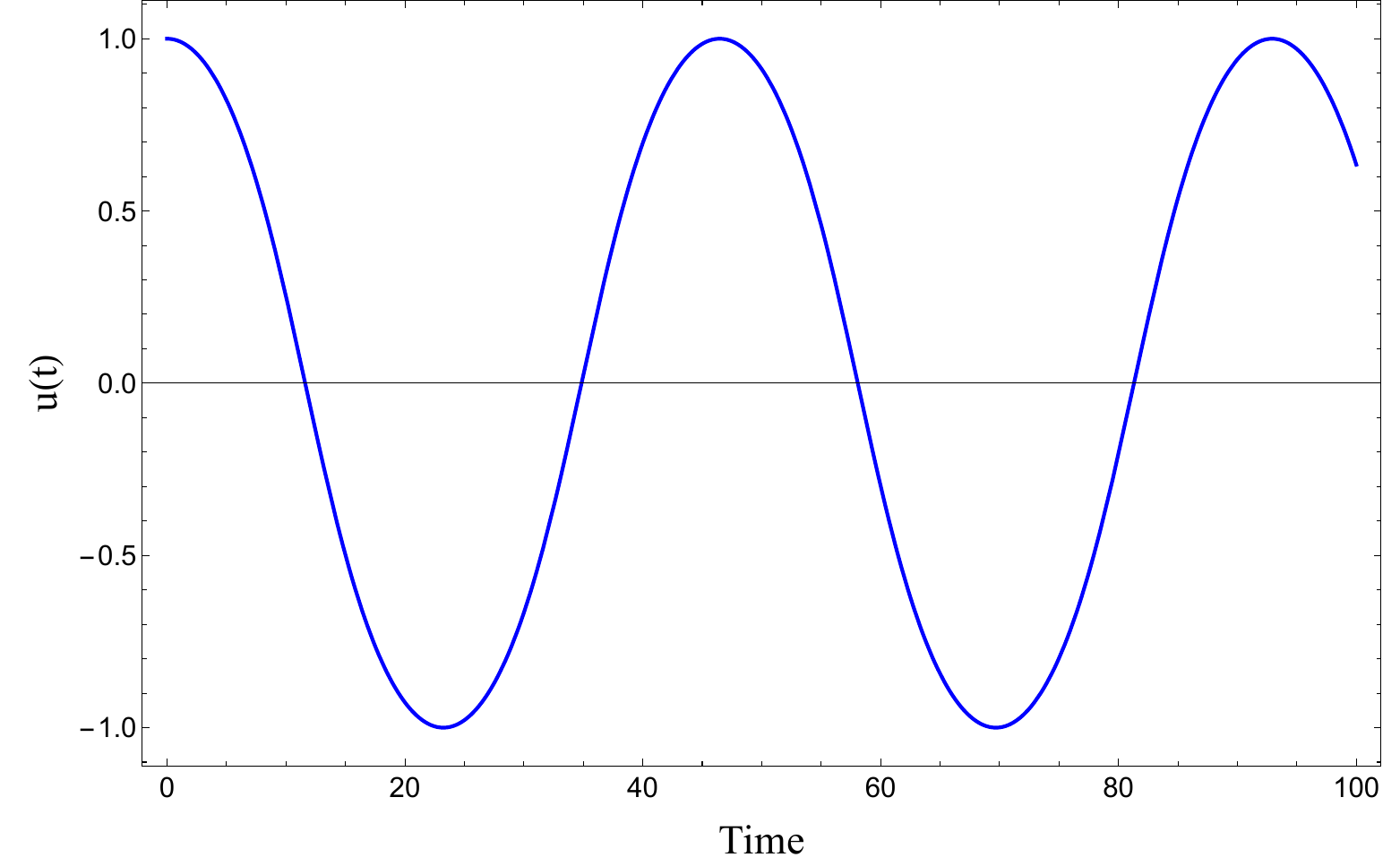}
		\caption{Evolution of radial parameter of wormhole.}
		\label{fig:um-BH-solution-2}
	\end{subfigure}%
	~ 
	~ 
	\caption{wormhole dynamics.}\label{fig:wormhole-dynamics-2.}
\end{figure}

\begin{figure}[bh]
	\centering
	\begin{subfigure}[b]{0.5\textwidth}
		\includegraphics[width=\textwidth]{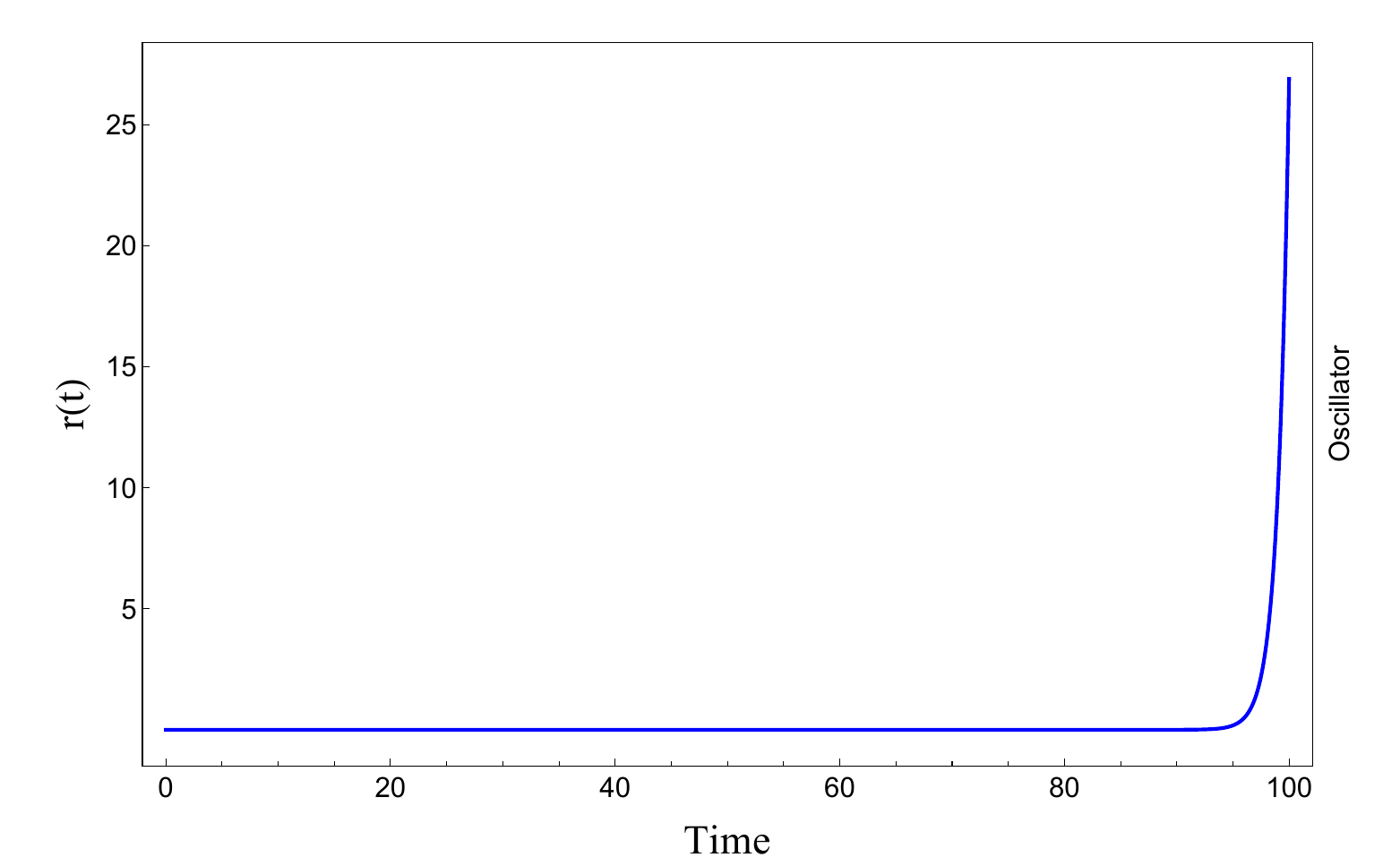}
		\caption{Evolution of oscillator's radial parameter. \qquad\qquad}
		\label{fig:um-oscillator-solution-2}
	\end{subfigure}%
	~ 
	\begin{subfigure}[b]{0.5\textwidth}
		\includegraphics[width=\textwidth]{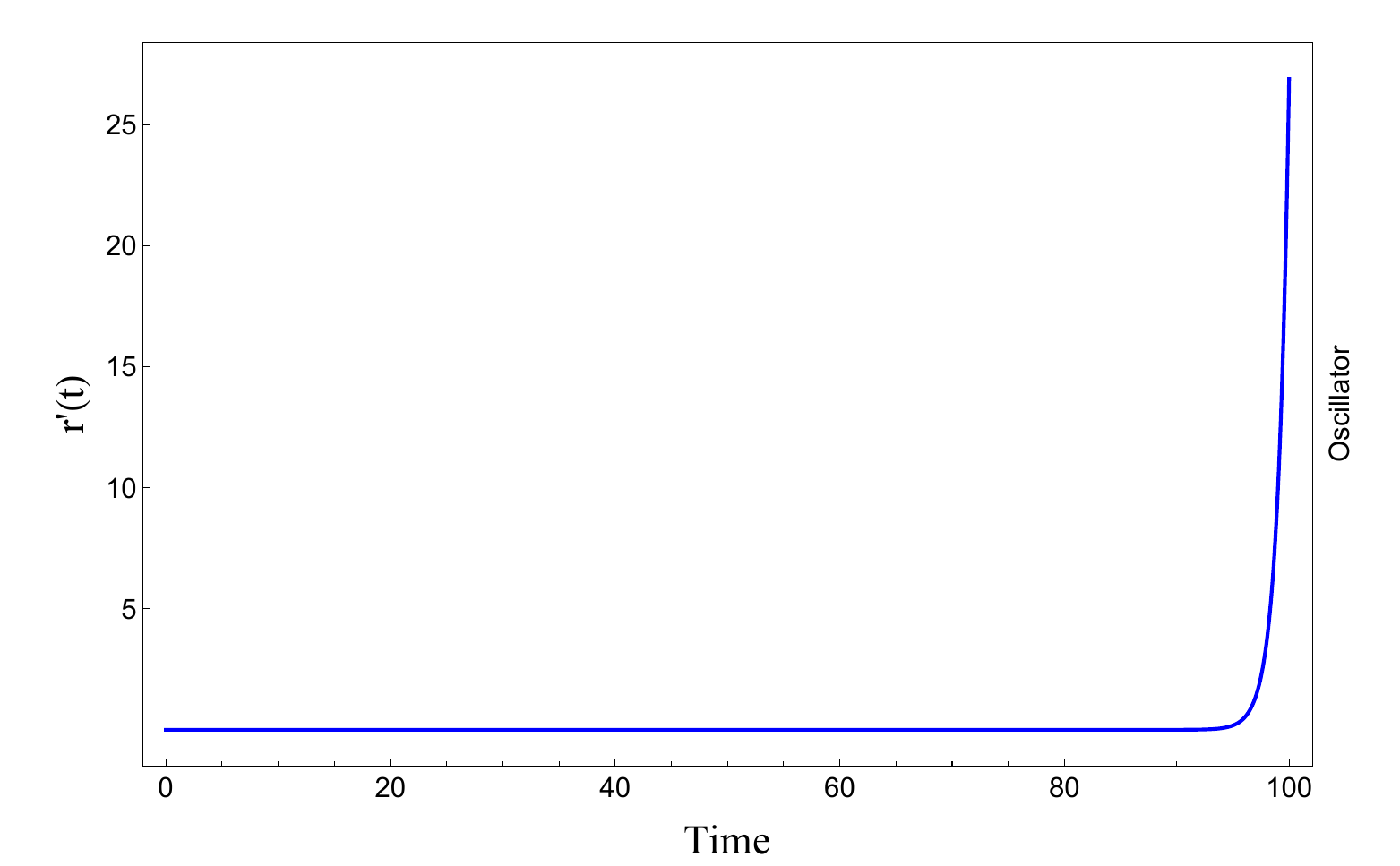}
		\caption{Evolution of oscillator's radial parameter's derivative.}
		\label{fig:uM-Oscillator-prime-solution-2}
	\end{subfigure}
	~ 
	\caption{Harmonic oscillator's dynamics.}\label{fig:Harmonic-oscillator's-dynamics-2.}
\end{figure}

From the plot presented in fig. (\ref{fig:wormhole-dynamics-2.}), in which the radial parameter of the wormhole is plotted as the function of time, we see that the wormhole here allows for the oscillatory behavior to occur in the ideal scenario. This result is as intuitively expected. There is no damping effect in the ideal oscillatory behavior thus the energy flow is possible without any loss. The situation when the oscillator is placed in the space-time is presented in fig. (\ref{fig:Harmonic-oscillator's-dynamics-2.}) and fig. (\ref{fig:um-oscillator-phasespace-solution-2}). 

The plots shown in fig. (\ref{fig:Harmonic-oscillator's-dynamics-2.}) are the dynamics of the harmonic oscillator when placed in the wormhole space-time. As we place an oscillator in the region, the gravity of the massive object (Schwarzschild black hole here) causes the oscillator to move towards the event horizon. 

The phase-space plot of the oscillator is shown in fig. (\ref{fig:um-oscillator-phasespace-solution-2}). Here we see that the transfer of the momentum (and hence the energy transfer) is smooth. As explained earlier, the dynamics of the oscillators happens in such a way that the velocity of the oscillator asymptotically increases such that the momentum in this scenario decreases as the event horizon is approached, therfore the energy is being transferred from the oscillator to the other side of the wormhole. 
\section{Conclusion}\label{conclusion}
We here present the wormhole formation in the very familiar Schwarzschild space-time. The wormhole is typically described as the structure that physically connects the two causally disconnected region of the space-time. This connection is assisted/formed by the presence of some massive object. 

In this article, we showed the proof of concept of the energy flow through the wormhole by placing a harmonic oscillator in the wormhole space-time. The harmonic oscillator is the natural choice of object since the objects are expressible in terms of  harmonic oscillators or their combination.

The Einstein Field Equations for the Schwarzschild geometry when a harmonic oscillator is place in it gave us two sets of solutions. In one outcome scenario, the dynamics of the oscillators occurred such that the velocity of the oscillator increased asymptotically. The momentum of the oscillator increased very rapidly as the consequence, we observe that as the oscillator approaches event horizon the momentum of the oscillator approaches infinity. This is shown in fig. (\ref{fig:um-oscillator-phasespace-solution-1}). We noticed an ideal oscillatory behavior suggesting that the energy can ideally flow without any loss which is presented in fig. (\ref{fig:wormhole-dynamics.}).

Another scenario which we obtained through EFEs solution showed oscillatory behavior of the oscillator without damping ideally as in the earlier solution. This once again suggested the energy flow without any loss in the ideal scenario which is shown in fig. (\ref{fig:wormhole-dynamics-2.}). The dynamics of the oscillators in this situation occurred such that the velocity of the oscillator increased asymptotically. The momentum of the oscillator however in this situation decreases as the event horizon is approached, therefore the energy is being transferred from the oscillator to the other side of the wormhole. This is shown in fig. (\ref{fig:um-oscillator-phasespace-solution-2}). 

The results show that the harmonic oscillators on the opposite sides of the wormhole transfer energy in between them through the wormhole. The oscillators are causally disconnected yet the gain in the energy of the oscillator in the first solution is occurring while the oscillator in the second solution is losing energy. The energy conservation therefore would follow
\begin{equation*}
	\Delta E_{osc-I}\leq \Delta E_{osc-II}~,
\end{equation*} 
where $\Delta E_{osc-I}$ and $\Delta E_{osc-II}$ are the change in the energy of the harmonic oscillator of solution-I and the change in the energy of the harmonic oscillator of solution-II respectively.

\end{document}